\documentclass[a4paper,12pt]{article}
\usepackage{latexsym}
\usepackage{amsmath}
\usepackage{amssymb}
\usepackage{amscd}
\usepackage{graphicx}

\newtheorem{theo}{Theorem}
\newtheorem{lem}[theo]{Lemma}

\newtheorem{prop}[theo]{Proposition}
\newtheorem{defi}[theo]{Definition}
\newcommand{\EQ}{\begin{equation}}
\newcommand{\EN}{\end{equation}}

\newcommand{\pr}{\indent{\em Proof. \ }}
\newcommand{\qed}{\hfill $\triangle$}

\newcommand{\Aut}{\mbox{\rm Aut}}

\newcommand{\wt}{\mbox{\rm wt}}

\newcommand{\by}{{\bf y}}
\newcommand{\bx}{{\bf x}}

\newcommand{\bv}{{\bf v}}

\newcommand{\bo}{{\bf 0}}
\newcommand{\bones}{\textbf{1}}

\newcommand{\F}{\mathbb{F}}

\newcommand{\cm}{C^{(m)}}

\newcommand{\cmp}{C^{[m]}}

\title{On a family of binary completely transitive codes with growing covering radius.\footnote{This work has been
partially supported by the Spanish MICINN grants MTM2009-08435;
the catalan grant 2009SGR1224 and also by the Russian fund of
fundamental researches 12 - 01 - 00905.\newline \indent $^1$J.
Rif\`{a} is with the Department of Information and Communications
Engineering, Universitat Aut\`{o}noma de Barcelona,
08193-Bellaterra, Spain. (email:~josep.rifa@uab.cat)
\newline \indent $^2$V. Zinoviev is with the A.A. Kharkevich
Institute for Problems of Information
Transmission, Russian Academy of Sciences, Bol'shoi Karetnyi
per. 19, GSP-4, Moscow, 127994, Russia (e-mail:\, zinov@iitp.ru).}}

\author{J. Rif\`{a}$^1$, V. A. Zinoviev$^2$}

\begin{document}

\maketitle

\begin{abstract}
A new family of binary linear
completely transitive (and, therefore, completely regular)
codes is constructed. The covering radius of these codes is growing with
the length of the code. In particular, for any integer
$\rho \geq 2$, there exist two codes with $d=3$, covering radius $\rho$
and length $\binom{4\,\rho}{2}$ and $\binom{4\,\rho+2}{2}$,
respectively. These new completely transitive codes induce, as
coset graphs, a family of distance-transitive graphs of growing
diameter.
\end{abstract}

\baselineskip=0.9\normalbaselineskip

\section{Introduction}
We use the standard notation $[n,k,d]$ for a binary
linear code $C$ of length $n$, dimension $k$ and minimum
distance $d$. The automorphism group $\Aut(C)$
coincides with the subgroup of the symmetric group
${\cal S}_n$ consisting of all $n!$ permutations of the
$n$ coordinate positions which send $C$ into itself.

Given any vector $\bv \in \F^n$, where $\F$ is the binary finite field,  its {\em distance} to
the code $C$ is $d(\bv,C)=\min_{\bx \in C}\{ d(\bv, \bx)\}$
and the {\em covering radius} of the code $C$ is
$\rho=\max_{\bv \in \F^n} \{d(\bv, C)\}$.

For a given code $C$ with covering radius $\rho=\rho(C)$ define
\[
C(i)~=~\{\bx \in
\F^{\,n}:\;d(\bx,C)=i\},\;\;i=1,2,...,\rho.
\]

\begin{defi}\label{de:1.1}
A code $C$ with covering radius $\rho=\rho(C)$ is
completely regular, if for all $l\geq 0$ every vector
$x \in C(l)$ has the same number $c_l$ of neighbors
in $C(l-1)$ and the same number $b_l$ of neighbors in
$C(l+1)$. Also, define $a_l = n-b_l-c_l$
and note that $c_0=b_\rho=0$. Define the intersection
array of $C$ as
$(b_0, \ldots, b_{\rho-1}; c_1,\ldots, c_{\rho})$.
\end{defi}

\begin{defi}\label{de:1.3}{\em \cite{sol,giu}}
A linear code $C$ with covering radius $\rho=\rho(C)$ and
automorphism group $\Aut(C)$ is completely transitive,
if the set of all cosets of $C$ is partitioned into
$\rho+1$ orbits under action of $\Aut(C)$, where for any
$\bx \in \F^n$ and $\varphi \in \Aut(C)$ the group acts
on a coset $\bx + C$ as
\[
\varphi (\bx + C)\;=\;\varphi(\bx) + C.
\]
\end{defi}

Let $\Gamma$ be a finite connected simple (i.e.,  undirected,
without loops and multiple edges) graph with diameter $\rho$. Let $d(\gamma, \delta)$
be the distance between two vertices $\gamma$ and $\delta$.
(i.e.,  a numbers of edges in the minimal path between $\gamma$
and $\delta$). Denote
\[
\Gamma_i(\gamma)~=~\{\delta\in\Gamma:~d(\gamma,\delta)=i\}.
\]
Two vertices $\gamma$ and $\delta$ from $\Gamma$ are {\em
neighbors} if $d(\gamma, \delta) = 1$. An {\em
automorphism} of a graph $\Gamma$ is a permutation $\pi$ of
the vertex set of $\Gamma$ such that, for all
$\gamma, \delta \in \Gamma$ we have $d(\gamma,\delta)=1$, if
and only if $d(\pi\gamma,\pi\delta)=1$. Let $\Gamma_i$ be a
subgraph of $\Gamma$ with the same vertices, where an edge
$(\gamma, \delta)$ is defined when the vertices $\gamma,
\delta$ are at distance $i$ in $\Gamma$. The graph $\Gamma$ is
called {\em primitive} if it is connected and and all $\Gamma_i$
~($i=1, \ldots, \rho$) are connected, and {\em imprimitive}
otherwise.

\begin{defi} {\em \cite{bro}}\label{14:de:1.4} A simple
connected graph $\Gamma$ is called {\em distance-regular},
if it is regular of valency $k$, and if for any two
vertices $\gamma, \delta \in \Gamma$ at distance $i$ apart,
there are precisely $c_i$ neighbors of $\delta$ in
$\Gamma_{i-1}(\gamma)$ and $b_i$ neighbors of $\delta$ in
$\Gamma_{i+1}(\gamma)$. Furthermore, this graph is called
{\em distance transitive}, if for any pair of vertices
$\gamma, \delta$ at distance $d(\gamma, \delta)$ there is
an automorphism $\pi\in\mbox{\rm Aut}(\Gamma)$ which
moves this pair to any other given pair $\gamma', \delta'$
of vertices at the same distance
$d(\gamma, \delta) = d(\gamma', \delta')$.
\end{defi}

Completely regular and completely transitive
codes are  classical subjects in algebraic
coding theory, which are closely connected
with graph theory, combinatorial designs and algebraic
combinatorics. Existence and enumeration of all such
codes are  open hard problems (see \cite{bro,del,neu} and
references there).

This paper is a natural continuation of our previous paper
\cite{rif}, where we describe a wide class of new binary
linear completely regular and completely transitive codes for
which the covering radius is growing with the length of the
code. The parameters of the main family of the codes depend
only on one integer parameter $m \geq 4$. The resulting code
$C$ has length $n=\binom{m}{2}$, dimension $k = n-m+1$, minimum distance $3$ and covering radius $\rho = \lfloor m/2 \rfloor$. A half of these
codes are non-antipodal and this implies (using \cite{bor}), that
the covering set $C(\rho)$ of $C$ is a coset of $C$. In these
cases the union $C \cup C(\rho)$ gives also a completely regular
and completely transitive code. Our purpose here is to
describe the resulting completely transitive codes.
We give, as a corollary of the constructed linear completely transitive
codes, an infinite family of distance-transitive coset graphs
with growing diameter.

\section{Preliminary results.}

Let $C$ be a binary linear completely regular code with covering
radius $\rho$ and intersection array\\
$(b_0, \ldots ,b_{\rho-1}; c_1, \ldots c_{\rho})$.
Let $\{D\}$ be the set of cosets of $C$. Define the graph
$\Gamma_C$ (which is called the {\em coset graph of $C$},
taking all cosets $D = C+ \bx$ as vertices, with two
vertices $\gamma = \gamma(D)$ and $\gamma' = \gamma(D')$
adjacent, if and only if the cosets $D$ and $D'$ contains
neighbor vertices, i.e.,  $\bv \in D$ and $\bv' \in D'$ with
distance $d(\bv, \bv') = 1$.

\begin{lem}\label{lem:2.2}{\em \cite{bro,rip}}.
Let $C$ be a linear completely regular code with covering
radius $\rho$ and intersection array
$(b_0, \ldots , b_{\rho-1}; c_1, \ldots c_{\rho})$
and let $\Gamma_C$ be the coset graph of $C$. Then $\Gamma_C$
is distance-regular of diameter $\rho$ with the same
intersection array. If $C$ is completely transitive, then
$\Gamma_C$ is distance-transitive.
\end{lem}

\begin{lem}\label{lem:2.3}{\em \cite{neu}}
Let $C$ be a completely regular code with covering radius
$\rho$ and intersection array
$(b_0, \ldots , b_{\rho-1}; c_1, \ldots c_{\rho})$.
Then $C(\rho)$ is a completely regular code too, with
inverse intersection array
$(b^r_0, \ldots , b^r_{\rho-1}; c^r_1, \ldots c^r_{\rho})$, i.e.,
\[
b^r_i~=~c_{\rho-i},~~~\mbox{and}~~~c^r_i~=~b_{\rho-i}.
\]
\end{lem}

The starting point for the results in this paper comes from~
\cite{rif}, where a specific class of combinatoric binary linear
codes was introduced. For a given natural number $m$ where
$m \geq 3$ denote by $E^m_2$ the set of all binary vectors of
length $m$ and weight $2$.

\begin{defi}\label{defi:3.1}
Let $H_m$ be the binary matrix of size $m \times m(m-1)/2$,
whose columns are exactly all the vectors from $E^m_2$ $($i.e.,
each vector from $E^m_2$ occurs once as a column of $H_m$).
Now define the binary linear code $C^{(m)}$ whose parity
check matrix is the matrix $H_m$.
\end{defi}

\begin{theo}{\em \cite{rif}}\label{theo:4.1}
Let $m$ be a natural number, $m \geq 3$. \\
(i) The binary linear $[n,k,d]$ code $C=C^{(m)}$ has
parameters:
\[
n~=~\binom{m}{\ell},~~k~=~n - m +
1,~~d~=~3,~~\rho~=~\lfloor \frac{m}{2} \rfloor.
\]
(ii) Code $C^{(m)}$ is completely transitive and, therefore,
completely regular. The intersection numbers of
$C^{(m)}$ for $i = 0, \ldots, \rho$ are:
\[
b_{i}~=~\binom{m-2i}{2},\;\;\;
c_{i}~=~\binom{2i}{2}.
\]
(iii) Code $C^{(m)}$ is antipodal if $m$ is odd
and non-antipodal if $m$ is even.
\end{theo}

\section{A new family of completely transitive codes}

Codes constructed in Theorem\cite{rif} for even $m$ are
non-antipodal and this implies (using \cite{bor}), that
the covering set $C(\rho)$ of $C$ is a coset of $C$. In these
cases the union $C \cup C(\rho)$ gives also a completely regular
and completely transitive code .

\begin{lem}\label{lem:3.2}
Let $C$ be a completely regular non-antipodal linear code with
$\bo\in C$. Then any coset $D=C+\bx$ of $C$ of weight $s$ is a
translate of $C(\rho)$ of weight $\rho-s$.
\end{lem}

\pr
Let $D=C+\bx$ be any coset of $C$ of weight $s$. We can take the
representative $\bx$ of weight $s$.

We want to show that $D=C(\rho)+\by$, where $y$ is a minimum
weight vector in $D$ and $\wt(\by)=\rho-s$. Since
$d(\bx,C)=s$, there exists a vector $\bv \in C(\rho)$, such
that $d(\bx,\bv)=\rho-s$ \cite{del}. Let $\by = \bv-\bx$.

Since for any $\bv\in C(\rho)$ we have $C+\textbf{1}=C+\bv=C(\rho)$,
we obtain
\[
D=C+\bx=C+(\bv+\by)=(C+\bv)+\by=C(\rho)+\by,
\]
which finishes the proof.
\qed

The next statement is very important for the results we obtain
in this paper.

\begin{theo}\label{theo:3.1}
Let $C$ be a non-antipodal code with $\bo \in C$, and let
$A=C \cup C(\rho)$.\\
$(i)$\;\; If $C$ is completely regular,
then the code $A$ is completely regular code too.\\
$(ii)$\; If $C$ is completely transitive, then $A$ is completely
transitive.
\end{theo}

\pr
$(i)$ Since $C$ is completely regular non-antipodal code, the set
$C(\rho)$ is a translate of $C$, i.e.,  $C(\rho)=C+\bf{1}$ \cite{bor}.
To show that $A$ is completely regular we check its
intersection array, denoted by $(b^a_0, \ldots , b^a_{\rho-1};
c^a_1, \ldots c^a_{\rho})$, where $\rho_a=\lfloor \rho/2 \rfloor$
is the covering radius of $A$. Recall (Lemma \ref{lem:2.3}) that
$C(\rho)$ is completely regular with intersection array
$(b^r_0, \ldots , b^r_{\rho-1}; c^r_1, \ldots c^r_{\rho})$,
where $b^r_i~=~c_{\rho-i}$ and $c^r_i~=~b_{\rho-i}$.

Now assume that $\bx \in A(s)$. Since $C(\rho)$ is a translate
of $C$, we have for $c^a_s$ and $b^a_s$ in any case, i.e.,
when $\bx \in C(s)$ or $\bx \in C(\rho)(s)$:
\[
\left\{\begin{array}{cllccccccc}
c^a_s&=&c_s, \;\;&b^a_s&=&b_s&\mbox{if }\;s&<&\lfloor \rho/2\rfloor,&\\
c^a_s&=&c_s+b_s,\;\;&b^a_s&=&0&\mbox{if }\;s&=&\rho/2,&\;\rho\;\;\mbox{even},\\
c^a_s&=&0,&b^a_s&=&b_s&\mbox{if }\;s&=&\lfloor \rho/2\rfloor,& \;\rho\;\;\mbox{odd}.
\end{array}
\right.
\]
Therefore, these numbers $c^a_s$ and $b^a_s$ do not depend on
the choice of the vector $\bx \in A(s)$. We conclude that
$A$ is completely regular.

$(ii)$ For the second statement we assume that $C$ is a (linear)
completely transitive code. Clearly, the code
$A=C\cup C(\rho)$ is a linear code with
covering radius $\rho_a=\lfloor \rho/2\rfloor$.
Now we have to show that for any two different vectors
$\bx,\bx' \in A(s)$,\;$1\leq s \leq \rho_a$
there is an automorphism (a permutation)
$\varphi \in \Aut(A)$ which transforms
the coset $B=A+\bx$ into the coset $B'=A+\bx'$.

We can assume that $\bx$ and $\bx'$ are representatives of minimum
weight $s$ in both cosets $B,B'$, respectively. Since $C$ is
completely transitive, the coset $D=C+\bx$ of $C$ can be
transformed into the coset $D'=C+\bx'$ by applying some
permutation $\varphi \in \Aut(C)$.  We claim that using the same
$\varphi$ we transform $B$ to $B'$. Note  that
$\Aut(C(\rho))=\Aut(C)$ (lemma \ref{lem:3.2}) and $\Aut(C)
\subseteq \Aut(A)$.


Since
\[
B=A+\bx=(C+\bx)\cup (C+\bf{1}+\bx),
\]
we obtain
\begin{eqnarray*}
\varphi(B)&=&\varphi(A+\bx)=\left(\varphi(C+\bx)\right)
\bigcup\left(\varphi(C+\bf{1}+\bx)\right)\\
&=&\left(C+\varphi(\bx)\right)\bigcup\left(C+\bf{1}+\varphi(\bx)\right)\\
&=&A+\varphi(\bx)=B'.
\end{eqnarray*}
\qed


Since for even $m$ the code $C^{(m)}$ is non-antipodal, its
covering set $C^{(m)}(\rho)$ is a translate of $C^{(m)}$
\cite{bor}. Hence, it makes sense to consider the new
(linear) code
\[
\cmp = C^{(m)} \cup
C^{(m)}(\rho).
\]

\begin{prop}\label{gm}
A generating matrix $G^{[m]}$ for code $\cmp$ is:
\[
G^{(m)}~=~\left[
\begin{array}{ccccc}
&I_{k-1}\,&|&\,H^T_{m-1}&\\\hline
&0 \ldots 0\,&|&\,1 \ldots 1&
\end{array}
\right]\,,
\]
where $H^T_{m-1}$ means the transpose matrix of $H_{m-1}$.
\end{prop}
\pr
Parity check matrix of $C^{(m)}$ can be written as
$\big( I_{m-1}|H_{m-1}\big)$ and so, the generating matrix
for $C^{(m)}$ is $\big( H^T_{m-1}|I_{n_{m-1}}\big)$, where
$n_{m-1}=k-1$ is the length of the code $C^{(m-1)}$.

Code $\cm$ is non-antipodal, so it does not contain the all
ones vector. We can add it to the generating matrix of $\cm$
to obtain $\cmp$ which gives the matrix $G^{[m]}$.
\qed

\bigskip

Next theorem gives important properties for code $\cmp$.

\begin{theo}\label{theo:4.2}
Let $m$ be even, $m \geq 6$ and let
$C^{[m]}=C^{(m)}\cup C^{(m)}(\rho)$. Then:\\
$(i)$\;\; Code $C^{[m]}$ is a linear completely regular linear $[n,k,d]$
code with parameters
\[
n~=~m(m-1)/2,~~k~=~n -
m + 2,~~d~=~3,~~\rho~=~
\lfloor m/4 \rfloor.
\]
$(ii)$\; The intersection numbers of $C^{[m]}$ for
$m \equiv 0 \pmod{4}$ and $\rho = m/4$ are
\[
b_{i}~=~\binom{m-2i}{2}\;\;\mbox{and}\;\;
c_{i}~=~\binom{2i}{2}\;\;\mbox{for}\;\;i
= 0,1, \ldots, \rho-1,\;\;
c_{\rho}~=~2\,\binom{2\rho}{2}\,,
\]
and, for $m \equiv 2 \pmod{4}$ and $\rho = (m-2)/4$, are
\[
b_{i}~=~\binom{m-2i}{2}\;\;\mbox{and}\;\;
c_{i}~=~\binom{2i}{2}\;\;\\\mbox{for}\;\;i
= 0,1, \ldots, \rho\,.
\]
$(iii)$ Code $C^{[m]}$ is completely transitive.
\end{theo}

\pr
This is a direct corollary of Theorem \ref{theo:3.1}.
\qed

\bigskip

We note that the extension of the code $C^{[m]}$ (i.e.,
adding one more overall parity checking position) is not
uniformly packed in the wide sense, and therefore, it is
not completely regular \cite{bro}.

\bigskip

Now we go over the structure of the automorphism group of the above codes.

\begin{lem}\label{lem:3.1}
Let $C$ be a binary completely regular code with $\bo\in C$ and let $\Aut(C)$ be
the automorphism group of $C$. If $C$ is not antipodal, then
$\Aut(C) = \Aut(C(\rho))$.
\end{lem}

\pr
Since $C$ is non-antipodal we have~\cite{bor} that $C(\rho) = C + \bf{1}$,
where $\textbf{1}=(1,1, \ldots, 1)$.

Now we have for any
$\varphi \in \Aut(C)$
\[
\varphi(C(\rho))=\varphi\left(C + \textbf{1}\right)
=\varphi(C)+ \varphi(\textbf{1})=C + \textbf{1}=C(\rho),
\]
which finishes the proof.
\qed

\begin{prop}
For the code $C^{(m)}$ we have $\Aut(C^{(m)})={\cal S}_m$.
\end{prop}
\pr
The automorphism groups of $C^{(m)}$ and its dual $C^{(m)\perp}$ coincides, so our argumentation is about the dual. It is easy to see that any permutation of the rows of $H_{m}$ gives the same code $C^{(m)\perp}$. A permutation of the rows in $H_{m}$ can be seen as a permutation on the columns of $H_{m}$, so an automorphism of $C^{(m)\perp}$. Hence, ${\cal S}_m \subseteq \Aut(C^{(m)\perp})$.

To show the reciprocal inclusion let us begin by the fact that all the codewords of weight $m-1$ in $C^{(m)\perp}$ are the rows of the matrix $H_m$. Indeed, the rows of matrix $H_m$ generate $C^{(m)\perp}$ and so, any codeword of weight $m-1$ in $C^{(m)\perp}$ is a linear combination of rows of $H_m$. If a linear combination $\bv$ of any $t$ rows is of weight $m-1$ we will have $\wt(\bv) = m-1=t·(m-1) - 2\binom{t}{2} \pmod{2^m-1}$ and so, $(t-1)(m-1)=t(t-1) \pmod{2^m-1}$. Hence, $t=1$ or $t=m-1$. In the first case $\bv$ is a row of $H_m$, but also in the second case (the sum of all rows in $H_m$ is the zero codeword, and so the sum of $m-1$ rows coincides with one row).

Now, we finish the proof of the statement. We want to prove that there are no more automorphisms in $\Aut(C^{(m)\perp})$ apart from these ones in ${\cal S}_m$. Assume $\phi\in \Aut(C^{(m)\perp})$. For all codewords $\bv\in C^{(m)\perp}$ of weight $m-1$  we have $\phi(\bv)\in C^{(m)\perp}$ is also of weight $m-1$. Hence, since the rows of $H_{m}$ are the only codewords of weight $m-1$ in $C^{(m)\perp}$ we have that $\phi$ is a permutation of the rows in $H_{m}$ and, as we said before, a permutation on the columns of $H_{m}$ and $\phi\in {\cal S}_{m}$.
\qed

\begin{prop} Let $m$ be even, $m\geq 6$.
\begin{enumerate}
\item \label{i} For $m>6$ we have  $\Aut(C^{[m]}) =\Aut(C^{(m)})$.
\item For $m=6$ we have $\Aut(C^{[m]})= GL_4(\F_2)$.
\end{enumerate}
\end{prop}
\pr
Generators matrix of $C^{(m)}$ and $C^{[m]}$ coincide, except
that the second one has one more independent row, the all ones
row $\bones$. Hence, all the coordinate permutations which fix
$C^{(m)}$ also fix $C^{[m]}$ and, so,
$\Aut(C^{(m)}) \subseteq \Aut(C^{[m]})$. For $m>6$, let
$\phi \in \Aut(C^{[m]})$ such that $\phi \notin  \Aut(C^{(m)})$.
Then, as the codewords of weight $3$ generate $C^{(m)}$
(straigforward from Proposition~\ref{gm}) there should exists a
codeword $\bv\in C^{(m)}$, of weight $3$  such that
$\phi(\bv)\in C^{(m)} + \bones$. Therefore, $\phi(\bv)$ should
be of weight 3, but this contradicts Theorem~\ref{theo:4.1}.
Hence, item~\ref{i} is proved.

For $m=6$ code $C^{[m]}$ is the Hamming code of length 15, and so its automorphism group is very well known. Indeed, it is the general linear group of degree 4.
\qed

\section{A new description of distance-transitive graphs}

Denote by $\Gamma^{(m)}$ (respectively, $\Gamma^{[m]}$) the
coset graph, obtained from the codes $C^{(m)}$ (respectively,
$C^{[m]}$) by Lemma \ref{lem:2.2}. From Theorems \ref{theo:4.1}
and \ref{theo:4.2} we obtain the following results, which leads
to new coset graphs.

\begin{theo}\label{theo:5.1}
$(i)$\;\; For any even $m \geq 6$ there exist two embedded double
covers $\Gamma^{(m)}$ and $\Gamma^{[m]}$ of complete graph
$K_n$, \,$n=\binom{m}{2}$, on $2^{m-1}$ and $2^{m-2}$ vertices,
respectively, and with covering radius $m/2$ and
$\lfloor m/4 \rfloor$, respectively.\\
$(ii)$\; The intersection arrays of graphs $\Gamma^{(m)}$
and $\Gamma^{[m]}$ are the same as the intersection arrays of
codes, given by Theorems \ref{theo:4.1} and \ref{theo:4.2}.\\
$(iii)$ Both graphs $\Gamma^{(m)}$ and $\Gamma^{[m]}$ are
distance transitive.\\
$(iv)$\; The graphs $\Gamma^{(m)}$ are imprimitive and the
graphs $\Gamma^{[m]}$ are primitive.\\
$(v)$\;\; The graph $\Gamma^{[m]}$ has eigenvalues:\\
for $m \equiv 2 \pmod{4}$
\[
\{\binom{m}{2} - 16i(\rho + 1 - i),\;i=0,1,\ldots,\rho\}
\]
and for $m \equiv 0 \pmod{4}$
\[
\{\binom{m}{2} - 8i(2\rho + 1 - i),\;i=0,1,\ldots,\rho\}\,.
\]
\end{theo}

The graph $\Gamma^{(m)}$ is well known. This graph can be
obtained from the even weight binary vectors of length $m$,
adjacent when their distance is $2$. It is the halved
$m$-cube and is a distance-transitive graph, uniquely defined
from its intersection array \cite[p. 264]{bro}. Since the graph
$\Gamma^{(m)}$ is antipodal, the graph $\Gamma^{[m]}$
(which has twice less vertices) can be seen as its folded graph,
obtained by collapsing antipodal pairs of vertices.

\end{document}